# Improved thermal lattice Boltzmann model for simulation of liquid-vapor phase change

Qing Li,[*] P. Zhou, and H. J. Yan

School of Energy Science and Engineering, Central South University, Changsha 410083, China

**Abstract**

In this paper, an improved thermal lattice Boltzmann (LB) model is proposed for simulating liquid-vapor phase change, which is aimed at improving an existing thermal LB model for liquid-vapor phase change [S. Gong and P. Cheng, Int. J. Heat Mass Transfer **55**, 4923 (2012)]. First, we emphasize that the replacement of $\nabla \cdot (\lambda \nabla T)/\rho c_V$ with $\nabla \cdot (\chi \nabla T)$ is an inappropriate treatment for diffuse interface modeling of liquid-vapor phase change. Furthermore, the error terms $\partial_{t_0}(T\mathbf{v}) + \nabla \cdot (T\mathbf{vv})$, which exist in the macroscopic temperature equation recovered from the standard thermal LB equation, are eliminated in the present model through a way that is consistent with the philosophy of the LB method. Moreover, the discrete effect of the source term is also eliminated in the present model. Numerical simulations are performed for droplet evaporation and bubble nucleation to validate the capability of the model for simulating liquid-vapor phase change. Numerical comparisons show that the replacement of $\nabla \cdot (\lambda \nabla T)/\rho c_V$ with $\nabla \cdot (\chi \nabla T)$ leads to significant numerical errors and the error terms in the recovered macroscopic temperature equation also result in considerable errors.

# I. INTRODUCTION

The lattice Boltzmann (LB) method, which originates from the lattice gas automata method [1], has been developed into an efficient numerical approach for a wide range of phenomena and processes in the past three decades [2-9]. The LB equation can be viewed as a special discrete solver for the kinetic Boltzmann equation with certain collision operator, such as the Bhatnagar-Gross-Krook (BGK) collision operator [3,10] and the multiple-relaxation-time (MRT) collision operator [11-16]. The fluid flow is simulated by tracking the evolution of the particle distribution function and then the distribution function is accumulated to obtain the macroscopic properties. The LB method is easy to parallelize and is far less costly in terms of data exchange owing to its explicit scheme and the local interactions.

In recent years, the LB simulations of liquid-vapor phase change have attracted much attention and three categories of thermal LB models have been developed for simulating liquid-vapor phase change. The first category is based on the phase-field multiphase LB method, such as the models developed by Dong *et al.* [17], Safari *et al.* [18,19], and Sun *et al.* [20]. In these models, the liquid-vapor interface is captured by solving an interface-capturing equation (e.g., the Cahn-Hilliard equation) and a source term is incorporated into the continuity equation or the interface-capturing equation to mimic the phase change. Hence the rate of the liquid-vapor phase change in these models is an artificial input.

The second category is based on the pseudopotential multiphase LB method, which is a very popular multiphase approach in the LB community [7]. In the pseudopotential multiphase LB method, the phase separation between different phases is achieved via an interparticle potential [21,22]. Therefore the liquid-vapor interface can naturally arise, deform, and migrate without using any interface-tracking or interface-capturing technique. The thermal multiphase LB models proposed by Zhang and Chen [23], Házi and Márkus [24,25], Biferale *et al.* [26], Gong and Cheng [27], Kamali *et al.* [28], and Li *et al.* [29] can be classified into this category. The third category is the multi-speed thermal LB method, which employs a single set of distribution functions like the standard isothermal LB method but utilizes more discrete velocities [30,31]. The equilibrium distribution function usually includes higher-order velocity terms so as to recover the energy equation. The thermal LB models presented by Gonnella *et al*. [32] and Gan *et al*. [33] for thermal liquid-vapor flows fall into this category.

In many of the aforementioned thermal multiphase LB models, a thermal LB equation is employed to recover a target temperature equation at the Navier-Stokes level. The target temperature equation is usually a convection-diffusion equation with a source term. Therefore a thermal LB equation with a source term was devised in these models. However, it has been widely found [34-37] that there exist error terms in the macroscopic equation recovered from the standard thermal LB equation, which should be treated using appropriate correction techniques. In addition, the temperature field can also be simulated by traditional numerical methods such as the finite-difference method. In Ref. [29], Li *et al*. devised a hybrid thermal LB model for liquid-vapor phase change, which employs a finite-difference scheme to solve the temperature equation.

Owing to the fact that many researchers prefer to use a thermal LB equation rather than a traditional numerical scheme, the thermal LB equation-based models are widely utilized in the literature for simulating liquid-vapor phase change. In particular, the thermal LB model proposed by Gong and Cheng [27] was recently used in some studies because of its simplicity, which results from the replacement of $\nabla \cdot (\lambda \nabla T)/\rho c_V$ with $\nabla \cdot (\chi \nabla T)$, where $\lambda$ is the thermal conductivity, $c_V$ is the specific heat at constant volume, and $\chi = \lambda/\rho c_V$ is the thermal diffusivity. Obviously, such a treatment is based on the assumption that the density $\rho$ is constant everywhere. Although the density in multiphase flows is constant in each single-phase region, it varies significantly within the liquid-vapor interface, which is usually a diffuse interface (around 4−5 lattices) in LB simulations.

In this work, we aim at presenting an improved thermal LB equation-based model for liquid-vapor phase change. The error terms $\partial_{t_0}(T\mathbf{v}) + \nabla \cdot (T\mathbf{vv})$, which arise from the standard thermal LB equation, are eliminated in a way that is consistent with the philosophy of the LB method. The discrete effect of the source term, which appears in previous thermal LB equation-based models for liquid-vapor phase change, is also eliminated in the present improved model. Numerical simulations show that the replacement of $\nabla \cdot (\lambda \nabla T)/\rho c_V$ with $\nabla \cdot (\chi \nabla T)$ leads to significant numerical errors. The rest of the present paper is organized as follows. The macroscopic temperature equation for liquid-vapor phase change and the thermal LB model proposed by Gong and Cheng are described in Sec. II. The improved thermal LB model is proposed in Sec. III. The numerical simulations and discussions are presented in Sec. IV. Finally, Sec. V concludes the present paper.

## II. Macroscopic temperature equation and the Gong-Cheng model

### A. The target temperature equation

Historically, the first thermal LB model for liquid-vapor phase change was proposed by Zhang and Chen [23]. In their work, the macroscopic energy equation was given by

$$\rho(\partial_t e + \mathbf{v} \cdot \nabla e) = -p \nabla \cdot \mathbf{v} + \nabla \cdot (\lambda \nabla T), \tag{1}$$

where $e$ is the internal energy, $c_V$ is the specific heat at constant volume, and $\lambda$ is the thermal conductivity. In 2009, Házi and Márkus [24] derived a target temperature equation from the local balance law for entropy [38]

$$\rho T \frac{\mathrm{D}s}{\mathrm{D}t} = \nabla \cdot (\lambda \nabla T), \tag{2}$$

where $s$ is the entropy and $\mathrm{D}(\bullet)/\mathrm{D}t = \partial_t (\bullet) + \mathbf{v} \cdot \nabla (\bullet)$ is the material derivative. The viscous heat dissipation has been neglected in Eq. (2). According to the thermodynamic relations of non-ideal gases, the following equation can be obtained:

$$\mathrm{d}s = \frac{c_V}{T} \mathrm{d}T + \left( \frac{\partial p_{\mathrm{EOS}}}{\partial T} \right)_V \mathrm{d}V, \tag{3}$$

where $p_{\mathrm{EOS}}$ is a non-ideal equation of state and $V = 1/\rho$ is the specific volume. The above equation is the first d$s$ equation in thermodynamics. According to Eq. (3) and the continuity equation $\mathrm{D}_t \rho = -\rho \nabla \cdot \mathbf{v}$, the following temperature equation can be derived from Eq. (2):

$$\rho c_V \left( \partial_t T + \mathbf{v} \cdot \nabla T \right) = \nabla \cdot (\lambda \nabla T) - T \left( \frac{\partial p_{\mathrm{EOS}}}{\partial T} \right)_\rho \nabla \cdot \mathbf{v}. \tag{4}$$

This equation can be found in Table 11.4-1 in Ref. [39]. For ideal gases ($p_{EOS} = \rho RT$), the last term on the right-hand side of Eq. (4) reduces to $p_{EOS} \nabla \cdot \mathbf{v}$. The above equation can also be written as follows:

$$\partial_t T + \mathbf{v} \cdot \nabla T = \frac{1}{\rho c_V} \nabla \cdot (\lambda \nabla T) - \frac{T}{\rho c_V} \left( \frac{\partial p_{EOS}}{\partial T} \right)_\rho \nabla \cdot \mathbf{v}. \tag{5}$$

In the literature, some other forms of the energy equation for non-ideal fluids can also be found [40-42]. For example, Onuki [41,42] established a general equation for the total energy density of non-ideal fluids (see Eq. (9) in Ref. [41]), which can be transformed to the following equation for the internal energy density (see Eq. (2.40) in Ref. [42]):

$$\partial_t \hat{e} + \nabla \cdot (\mathbf{v}\hat{e}) = \nabla \cdot (\lambda \nabla T) - (\mathbf{\Pi} - \boldsymbol{\sigma}) : \nabla \mathbf{v}, \tag{6}$$

where $\hat{e} = \rho e$ is the internal energy density ($e$ is the internal energy of non-ideal fluids), $\boldsymbol{\sigma}$ is the dissipative stress tensor, and $\mathbf{\Pi} = p_{EOS}\mathbf{I} + \mathbf{T}$ is the nonviscous stress, in which $p_{EOS}$ is the non-ideal equation of state, $\mathbf{I}$ is the unit tensor, and $\mathbf{T}$ is the contribution to the pressure tensor depending on density gradients [32]. Using the continuity equation and $\mathbf{I} : \nabla \mathbf{v} = \nabla \cdot \mathbf{v}$, the following equation can be derived from Eq. (6):

$$\rho \frac{\mathrm{D}e}{\mathrm{D}t} = \nabla \cdot (\lambda \nabla T) - p_{EOS} \nabla \cdot \mathbf{v} - (\mathbf{T} - \boldsymbol{\sigma}) : \nabla \mathbf{v}. \tag{7}$$

According to thermodynamics, the relationship between the internal energy and the entropy is given by

$$\mathrm{d}e = T\mathrm{d}s - p_{EOS}\mathrm{d}V, \tag{8}$$

where $V = 1/\rho$. Using Eq. (8), the internal energy equation (7) can be transformed to

$$\rho \left[ T \frac{\mathrm{D}s}{\mathrm{D}t} - p_{EOS} \frac{\mathrm{D}}{\mathrm{D}t}\left(\frac{1}{\rho}\right) \right] = \nabla \cdot (\lambda \nabla T) - p_{EOS} \nabla \cdot \mathbf{v} - (\mathbf{T} - \boldsymbol{\sigma}) : \nabla \mathbf{v}. \tag{9}$$

Substituting the continuity equation $\mathrm{D}\rho/\mathrm{D}t = -\rho \nabla \cdot \mathbf{v}$ into Eq. (9) yields

$$\rho T \frac{\mathrm{D}s}{\mathrm{D}t} = \nabla \cdot (\lambda \nabla T) - (\mathbf{T} - \boldsymbol{\sigma}) : \nabla \mathbf{v}. \tag{10}$$

The term $\boldsymbol{\sigma} : \nabla \mathbf{v}$ represents the viscous heat dissipation. Comparing Eq. (2) with Eq. (10), we can see that these two equations are basically consistent except that $(\mathbf{T} - \boldsymbol{\sigma}) : \nabla \mathbf{v}$ is neglected in Eq. (2).

**B. The Chapman-Enskog analysis of the Gong-Cheng model**

For simplicity, Gong and Cheng [27] replaced $\nabla \cdot (\lambda \nabla T)/\rho c_V$ in Eq. (5) with $\nabla \cdot (\chi \nabla T)$, where $\chi = \lambda/\rho c_V$ is the thermal diffusivity. Then they established the following temperature equation:

$$\partial_t T + \nabla \cdot (\mathbf{v} T) = \nabla \cdot (\chi \nabla T) + T\left[1 - \frac{1}{\rho c_V}\left(\frac{\partial p_{\text{EOS}}}{\partial T}\right)_\rho\right]\nabla \cdot \mathbf{v}. \tag{11}$$

The corresponding thermal LB equation for Eq. (11) was given by [27]

$$g_\alpha(\mathbf{x} + \mathbf{e}_\alpha \delta_t, t + \delta_t) - g_\alpha(\mathbf{x}, t) = -\frac{1}{\tau_g}\left(g_\alpha - g_\alpha^{eq}\right) + \delta_t G_\alpha, \tag{12}$$

where $g_\alpha$ is the temperature distribution function, $\mathbf{e}_\alpha$ is the discrete velocity in the $\alpha$ th direction, $\tau_g$ is non-dimensional relaxation time for the temperature field, and the source term $G_\alpha = \omega_\alpha \phi$, in which $\phi$ represents the second term on the right-hand side of Eq. (11), namely [27]

$$\phi = T\left[1 - \frac{1}{\rho c_V}\left(\frac{\partial p_{\text{EOS}}}{\partial T}\right)_\rho\right]\nabla \cdot \mathbf{v}. \tag{13}$$

The equilibrium temperature distribution function $g_\alpha^{eq}$ was given by

$$g_\alpha^{eq} = \omega_\alpha T\left[1 + \frac{\mathbf{e}_\alpha \cdot \mathbf{v}}{c_s^2} + \frac{\mathbf{v}\mathbf{v}:(\mathbf{e}_\alpha \mathbf{e}_\alpha - c_s^2 \mathbf{I})}{2c_s^4}\right], \tag{14}$$

where $\mathbf{I}$ is the unit tensor, $c_s = c/\sqrt{3}$ is the lattice sound speed, and $\omega_\alpha$ are the weights, which are given by (for the D2Q9 lattice): $\omega_0 = 4/9$, $\omega_{1-4} = 1/9$, and $\omega_{5-8} = 1/36$.

The macroscopic equation recovered from Eq. (12) can be derived through the Chapman-Enskog analysis, which can be implemented by introducing the following multi-scale expansions:

$$\partial_t = \partial_{t_0} + \delta_t \partial_{t_1}, \quad g_\alpha = g_\alpha^{eq} + \delta_t g_\alpha^{(1)} + \delta_t^2 g_\alpha^{(2)}, \tag{15}$$

where $t_0$ and $t_1$ are two different time scales, and $\delta_t$ serves as the expansion parameter [43]. The Taylor series expansion of Eq. (12) yields

$$\delta_t (\partial_t + \mathbf{e}_\alpha \cdot \nabla) g_\alpha + \frac{\delta_t^2}{2}(\partial_t + \mathbf{e}_\alpha \cdot \nabla)^2 g_\alpha + \cdots = -\frac{1}{\tau_g}\left(g_\alpha - g_\alpha^{eq}\right) + \delta_t G_\alpha. \tag{16}$$

With the help of Eq. (15), Eq. (16) can be rewritten in the consecutive orders of $\delta_t$ as follows:

$$O(\delta_t): \left(\partial_{t_0} + \mathbf{e}_\alpha \cdot \nabla\right)g_\alpha^{eq} = -\frac{1}{\tau_g} g_\alpha^{(1)} + G_\alpha, \tag{17}$$

$$O\left(\delta_t^2\right): \partial_{t_1} g_\alpha^{eq} + \left(\partial_{t_0} + \mathbf{e}_\alpha \cdot \nabla\right) g_\alpha^{(1)} + \frac{1}{2}\left(\partial_{t_0} + \mathbf{e}_\alpha \cdot \nabla\right)^2 g_\alpha^{eq} = -\frac{1}{\tau_g} g_\alpha^{(2)}. \tag{18}$$

Substituting Eq. (17) into Eq. (18) leads to

$$\partial_{t_1} g_\alpha^{eq} + \left(\partial_{t_0} + \mathbf{e}_\alpha \cdot \nabla\right)\left(1 - \frac{1}{2\tau_g}\right) g_\alpha^{(1)} + \frac{1}{2}\left(\partial_{t_0} + \mathbf{e}_\alpha \cdot \nabla\right) G_\alpha = -\frac{1}{\tau_g} g_\alpha^{(2)}. \tag{19}$$

Taking the summations of Eqs. (17) and (19), the following equations can be obtained, respectively:

$$\partial_{t_0} T + \nabla \cdot (\mathbf{v} T) = \phi, \tag{20}$$

$$\partial_{t_1} T + \nabla \cdot \left(1 - \frac{1}{2\tau_g}\right)\left(\sum_\alpha \mathbf{e}_\alpha g_\alpha^{(1)}\right) + \frac{1}{2} \partial_{t_0} \phi = 0. \tag{21}$$

In the above derivations, the relations $\sum_\alpha g_\alpha^{(1)} = \sum_\alpha g_\alpha^{(2)} = 0$, $\sum_\alpha G_\alpha = \phi$, and $\sum_\alpha \mathbf{e}_\alpha G_\alpha = 0$ have been used. From Eq. (17) we can obtain

$$\sum_\alpha \mathbf{e}_\alpha g_\alpha^{(1)} = -\tau_g \left[\partial_{t_0}\left(\sum_\alpha \mathbf{e}_\alpha g_\alpha^{eq}\right) + \nabla \cdot \left(\sum_\alpha \mathbf{e}_\alpha \mathbf{e}_\alpha g_\alpha^{eq}\right)\right]. \tag{22}$$

With the aid of Eq. (14), we have

$$\sum_\alpha \mathbf{e}_\alpha g_\alpha^{(1)} = -\tau_g \left[\partial_{t_0}(T\mathbf{v}) + \nabla \cdot (T\mathbf{v}\mathbf{v}) + c_s^2 \nabla T\right]. \tag{23}$$

Substituting Eq. (23) into Eq. (21) gives

$$\partial_{t_1} T = \nabla \cdot \left\{(\tau_g - 0.5)\left[\partial_{t_0}(T\mathbf{v}) + \nabla \cdot (T\mathbf{v}\mathbf{v}) + c_s^2 \nabla T\right]\right\} - \frac{1}{2} \partial_{t_0} \phi. \tag{24}$$

Combining Eq. (20) with Eq. (24) through $\partial_t = \partial_{t_0} + \delta_t \partial_{t_1}$, we can obtain

$$\partial_t T + \nabla \cdot (\mathbf{v} T) = \nabla \cdot (\chi \nabla T) + \phi + \underline{\nabla \cdot \left\{(\tau_g - 0.5)\delta_t \left[\partial_{t_0}(T\mathbf{v}) + \nabla \cdot (T\mathbf{v}\mathbf{v})\right]\right\} - \frac{\delta_t}{2} \partial_{t_0} \phi}, \tag{25}$$

where $\chi = (\tau_g - 0.5) c_s^2 \delta_t$. The above equation is the macroscopic temperature equation recovered from Eq. (12). The underlined terms in Eq. (25) are additional (error) terms, which also appear in some other thermal LB equation-based models for liquid-vapor phase change. Among these error terms, the error terms $\partial_{t_0}(T\mathbf{v}) + \nabla \cdot (T\mathbf{v}\mathbf{v})$ result from $\sum_\alpha \mathbf{e}_\alpha g_\alpha^{(1)}$, while the last term on the right-hand side of Eq. (25) is caused by the discrete effect of the source term, which can be seen from Eqs. (19) and (21).

***Remark 1***. The replacement of $\nabla \cdot (\lambda \nabla T)/\rho c_V$ with $\nabla \cdot (\chi \nabla T)$ is an inappropriate treatment for multiphase flows. In fact, such a treatment requires that the following term can be neglected:

$$\varphi = \frac{\nabla \cdot (\lambda \nabla T)}{\rho c_V} - \nabla \cdot \left( \frac{\lambda \nabla T}{\rho c_V} \right) \equiv \frac{(\lambda \nabla T) \cdot \nabla (\rho c_V)}{(\rho c_V)^2}. \quad (26)$$

For single-phase incompressible flows, the aforementioned replacement is applicable since the density variation is very small. For multiphase flows, the density varies significantly within the liquid-vapor interface, which usually has a thickness of 4−5 lattices in the LB simulations of multiphase flows. Therefore the term given by Eq. (26) cannot be neglected at the liquid-vapor interface. Some researchers [44] found that under certain conditions $\varphi$ is small in comparison with the source term $\phi$ given by Eq. (13). In fact, not only $\varphi$ but also the thermal conductivity term $\nabla \cdot (\lambda \nabla T)/\rho c_V$ can be small as compared with the term $\phi$ in Eq. (13), but it does not mean that $\varphi$ or the thermal conductivity term in the temperature equation can be dropped. The comparison should be made between $\varphi$ and the thermal conductivity term instead of comparing $\varphi$ with the source term $\phi$ in Eq. (13), because it arises from the replacement of $\nabla \cdot (\lambda \nabla T)/\rho c_V$ with $\nabla \cdot (\chi \nabla T)$.

*Remark 2*. The error terms in the recovered macroscopic temperature equation are usually very small for sing-phase incompressible flows. Nevertheless, they may result in considerable errors for multiphase flows. For example, the error term $\partial_{t_0}(T\mathbf{v})$ can be split into $\partial_{t_0}(T\mathbf{v}) = \mathbf{v}\partial_{t_0}T + T\partial_{t_0}\mathbf{v}$, in which $\partial_{t_0}\mathbf{v}$ is given as follows according to the Chapman-Enskog analysis of the LB equation for the flow field [36]:

$$\partial_{t_0}\mathbf{v} = -\mathbf{v} \cdot \nabla \mathbf{v} + \frac{1}{\rho}\left[\mathbf{F} - \nabla(\rho c_s^2)\right], \quad (27)$$

where $\mathbf{F}$ is the force acting on the system. Obviously, $\mathbf{F}/\rho$ and $\nabla \rho/\rho$ are non-negligible within the liquid-vapor interface for multiphase flows.

### III. Improved thermal LB model

#### A. Theoretical analysis based on the BGK collision operator

The improved thermal LB model will be constructed based on the MRT collision operator. Before presenting the improved model, we would like to provide some analyses about removing the error terms in Eq. (25) within the framework of the BGK collision operator, which may be useful for general readers to better understand the improved thermal LB model in the next subsection. The target temperature equation given by Eq. (5) can be rewritten as follows [25]:

$$\partial_t T + \nabla \cdot (\mathbf{v}T) = \nabla \cdot (k\nabla T) + \underline{\frac{1}{\rho c_V}\nabla \cdot (\lambda \nabla T) - \nabla \cdot (k\nabla T) + T\left[1 - \frac{1}{\rho c_V}\left(\frac{\partial p_{\text{EOS}}}{\partial T}\right)_\rho\right]\nabla \cdot \mathbf{v}}. \quad (28)$$

The source term $\phi$ is now given by the underlined terms in Eq. (28).

According to Eqs. (22) and (23), the error term $\nabla \cdot (T\mathbf{v}\mathbf{v})$ in Eq. (25) can be removed by dropping the second-order velocity terms in $g_\alpha^{eq}$, and then $g_\alpha^{eq}$ becomes

$$g_\alpha^{eq} = \omega_\alpha T\left(1 + \frac{\mathbf{e}_\alpha \cdot \mathbf{v}}{c_s^2}\right). \quad (29)$$

Meanwhile, the error term $\partial_{t_0}(T\mathbf{v})$ in Eq. (25) can be eliminated by adding a correction term to the thermal LB equation

$$g_\alpha(\mathbf{x} + \mathbf{e}_\alpha \delta_t, t + \delta_t) - g_\alpha(\mathbf{x}, t) = -\frac{1}{\tau_g}(g_\alpha - g_\alpha^{eq}) + \delta_t G_\alpha + \delta_t C_\alpha, \quad (30)$$

where the correction term $C_\alpha$ is given by

$$C_\alpha = \left(1 - \frac{1}{2\tau_g}\right)\frac{\omega_\alpha \mathbf{e}_\alpha \cdot \partial_t (T\mathbf{v})}{c_s^2}, \quad (31)$$

which satisfies $\sum_\alpha C_\alpha = 0$ and $\sum_\alpha \mathbf{e}_\alpha C_\alpha = (1 - 0.5/\tau_g)\partial_t(T\mathbf{v})$.

Theoretically, to remove the discrete effect of the source term, namely the error term $\partial_{t_0}\phi$ in Eq. (25), the source term $G_\alpha$ in Eq. (30) should also contain the coefficient $(1 - 0.5/\tau_g)$ in the correction term given by Eq. (31), which has been extensively demonstrated in the literature when a forcing or source term is incorporated into the LB equation [45,46]. However, when this coefficient is placed in front of the source term, the temperature should be calculated by $T = \sum_\alpha g_\alpha + 0.5\delta_t \sum_\alpha G_\alpha$. Since $G_\alpha = \omega_\alpha \phi$, in which $\phi$ contains $\nabla \cdot (\lambda \nabla T)$, the calculation of the temperature will become implicit and iterations will be required.

Hence another treatment is considered. If we retain the definition of the temperature $T = \sum_\alpha g_\alpha$, the source term should take the form of $0.5\delta_t[G_\alpha(\mathbf{x} + \mathbf{e}_\alpha \delta_t, t + \delta_t) + G_\alpha(\mathbf{x}, t)]$ so as to remove the discrete effect of the source term. Fortunately, we have $\sum_\alpha \mathbf{e}_\alpha G_\alpha = 0$, hence the term $\mathbf{e}_\alpha \cdot \nabla G_\alpha$ in Eq. (19) does not affect the summation of Eq. (19). Therefore the source term can take the following form:

$$\frac{\delta_t}{2}\left[G_\alpha(\mathbf{x}, t+\delta_t) + G_\alpha(\mathbf{x}, t)\right] \approx \delta_t\left(1 + \frac{\delta_t}{2}\partial_t\right)G_\alpha(\mathbf{x}, t). \tag{32}$$

Then the thermal LB equation becomes

$$g_\alpha(\mathbf{x} + \mathbf{e}_\alpha \delta_t, t+\delta_t) - g_\alpha(\mathbf{x}, t) = -\frac{1}{\tau_g}(g_\alpha - g_\alpha^{eq}) + \delta_t C_\alpha + \delta_t\left(G_\alpha + \frac{\delta_t}{2}\partial_t G_\alpha\right). \tag{33}$$

The Chapman-Enskog analysis can also be performed for Eq. (33). Using the multi-scale expansions given by Eq. (15), the correction term $C_\alpha$ should be expanded as $C_\alpha = C_{\alpha 0} + \delta_t C_{\alpha 1}$ since it contains $\partial_t(T\mathbf{v})$. Then the following equations can be obtained:

$$O(\delta_t): \left(\partial_{t_0} + \mathbf{e}_\alpha \cdot \nabla\right)g_\alpha^{eq} = -\frac{1}{\tau_g}g_\alpha^{(1)} + C_{\alpha 0} + G_\alpha, \tag{34}$$

$$O(\delta_t^2): \partial_{t_1}g_\alpha^{eq} + \left(\partial_{t_0} + \mathbf{e}_\alpha \cdot \nabla\right)g_\alpha^{(1)} + \frac{1}{2}\left(\partial_{t_0} + \mathbf{e}_\alpha \cdot \nabla\right)^2 g_\alpha^{eq} = -\frac{1}{\tau_g}g_\alpha^{(2)} + C_{\alpha 1} + \frac{1}{2}\partial_{t_0}G_\alpha, \tag{35}$$

Substituting Eq. (34) into Eq. (35) yields

$$\partial_{t_1}g_\alpha^{eq} + \left(\partial_{t_0} + \mathbf{e}_\alpha \cdot \nabla\right)\left(1 - \frac{1}{2\tau_g}\right)g_\alpha^{(1)} + \frac{1}{2}(\mathbf{e}_\alpha \cdot \nabla)G_\alpha + \frac{1}{2}\left(\partial_{t_0} + \mathbf{e}_\alpha \cdot \nabla\right)C_{\alpha 0} = -\frac{1}{\tau_g}g_\alpha^{(2)} + C_{\alpha 1}. \tag{36}$$

Note that the last term on the right-hand side of Eq. (35) has been used to eliminate the same term generated on the left-hand side of Eq. (36). The summations of Eqs. (34) and (36) lead to, respectively

$$\partial_{t_0}T + \nabla \cdot (\mathbf{v}T) = \phi, \tag{37}$$

$$\partial_{t_1}T + \nabla \cdot \left(1 - \frac{1}{2\tau_g}\right)\left(\sum_\alpha \mathbf{e}_\alpha g_\alpha^{(1)}\right) + \frac{1}{2}\nabla \cdot \left(\sum_\alpha \mathbf{e}_\alpha C_{\alpha 0}\right) = 0. \tag{38}$$

In the above derivations, the relations $\sum_\alpha G_\alpha = \phi$, $\sum_\alpha \mathbf{e}_\alpha G_\alpha = 0$, and $\sum_\alpha C_{\alpha 0} = \sum_\alpha C_{\alpha 1} = 0$ have been used. From Eq. (34) we can obtain

$$\sum_\alpha \mathbf{e}_\alpha g_\alpha^{(1)} = -\tau_g\left[\partial_{t_0}\left(\sum_\alpha \mathbf{e}_\alpha g_\alpha^{eq}\right) + \nabla \cdot \left(\sum_\alpha \mathbf{e}_\alpha \mathbf{e}_\alpha g_\alpha^{eq}\right) - \sum_\alpha \mathbf{e}_\alpha C_{\alpha 0}\right]. \tag{39}$$

Using Eq. (29), we have

$$\partial_{t_1}T = \nabla \cdot \left\{(\tau_g - 0.5)\left[\partial_{t_0}(T\mathbf{v}) + c_s^2 \nabla T\right]\right\} - \nabla \cdot \left(\tau_g \sum_\alpha \mathbf{e}_\alpha C_{\alpha 0}\right). \tag{40}$$

Since $\sum_\alpha \mathbf{e}_\alpha C_{\alpha 0} = (1 - 0.5/\tau_g)\partial_{t_0}(T\mathbf{v})$, the error term $\partial_{t_0}(T\mathbf{v})$ in Eq. (40) can be eliminated. Then the target temperature equation can be correctly recovered as follows:

$$\partial_t T + \nabla \cdot (\mathbf{v}T) = \nabla \cdot (k\nabla T) + \phi, \tag{41}$$

where $k = (\tau_g - 0.5)c_s^2 \delta_t$ and $\phi$ denotes the underlined terms in Eq. (28).

It can be found that the following treatments have been employed in the above analyses. First, a correct target temperature equation is adopted. Second, the error terms $\partial_{t_0}(T\mathbf{v}) + \nabla \cdot (T\mathbf{vv})$ are removed by dropping the second-order velocity terms in the equilibrium temperature distribution function and adding a correction term to the thermal LB equation. Furthermore, the discrete effect of the source term is eliminated by incorporating an additional term into the thermal LB equation.

## B. The improved thermal MRT-LB model

In this subsection, the improved thermal LB model is presented based on the MRT collision operator. Using the MRT collision operator, the thermal LB equation can be written as follows:

$$g_\alpha(\mathbf{x} + \mathbf{e}_\alpha \delta_t, t + \delta_t) = g_\alpha(\mathbf{x}, t) - \bar{\Lambda}_{\alpha\beta}(g_\beta - g_\beta^{eq})|_{(\mathbf{x},t)} + \delta_t S'_\alpha(\mathbf{x}, t), \tag{42}$$

where $S'_\alpha$ is the source term in the discrete velocity space and $\bar{\Lambda}_{\alpha\beta} = (\mathbf{M}^{-1}\mathbf{\Lambda}\mathbf{M})_{\alpha\beta}$ is the collision matrix [11,47], in which $\mathbf{M}$ is an orthogonal transformation matrix and $\mathbf{\Lambda}$ is a diagonal matrix given by (for the D2Q9 lattice)

$$\mathbf{\Lambda} = \text{diag}(s_0, s_1, s_2, s_3, s_4, s_5, s_6, s_7, s_8). \tag{43}$$

Through the transformation matrix $\mathbf{M}$, the temperature distribution function $g_\alpha$ and its equilibrium distribution $g_\alpha^{eq}$ can be projected onto the moment space via $\mathbf{m} = \mathbf{Mg}$ and $\mathbf{m}^{eq} = \mathbf{Mg}^{eq}$, respectively, where $\mathbf{g} = (g_0, g_1, \cdots, g_8)^\mathrm{T}$ and $\mathbf{g}^{eq} = (g_0^{eq}, g_1^{eq}, \cdots, g_8^{eq})^\mathrm{T}$.

The second-order velocity terms in the equilibrium distribution function $g_\alpha^{eq}$ should be dropped to remove the error term $\nabla \cdot (T\mathbf{vv})$. The equilibria $\mathbf{m}^{eq}$ that correspond to Eq. (29) are given by

$$\mathbf{m}^{eq} = T(1, -2, 1, v_x, -v_x, v_y, -v_y, 0, 0)^\mathrm{T}. \tag{44}$$

The right-hand side of Eq. (42) can be implemented in the moment space as follows:

$$\mathbf{m}^* = \mathbf{m} - \mathbf{\Lambda}(\mathbf{m} - \mathbf{m}^{eq}) + \delta_t \mathbf{S}, \tag{45}$$

where $\mathbf{m}^* = (m_0^*, m_1^*, \cdots, m_8^*)^{\mathrm{T}}$ and $\mathbf{S}$ is the source term in the moment space. We are not concerned about the detailed form of $S'_\alpha$ in the discrete velocity space since the source term $\mathbf{S}$ can be directly specified in the moment space. For the present improved model, the source term $\mathbf{S}$ is given by

$$\mathbf{S} = (S_0, 0, 0, 0, 0, 0, 0, 0, 0)^{\mathrm{T}}, \tag{46}$$

where $S_0 = \phi + 0.5\delta_t \partial_t \phi$. As discussed in the previous subsection, the additional term $0.5\delta_t \partial_t \phi$ is used to eliminate the discrete effect of the source term. With Eq. (45), the streaming process is given by

$$g_\alpha(\mathbf{x} + \mathbf{e}_\alpha \delta_t, t + \delta_t) = g_\alpha^*(\mathbf{x}, t), \tag{47}$$

where $\mathbf{g}^* = \mathbf{M}^{-1}\mathbf{m}^*$. With the above treatments, it can be found that the error term $\nabla \cdot (T\mathbf{v}\mathbf{v})$ and the discrete effect of the source term have been eliminated. However, the error term $\partial_{t_0}(T\mathbf{v})$ still exists, which can be seen from the Chapman-Enskog analysis given in the Appendix.

Similar to the treatment based on the BGK collision operator, the error term $\partial_{t_0}(T\mathbf{v})$ can be eliminated by adding correction terms to the collision processes of $m_3$ and $m_5$, respectively

$$m_{3,\text{new}}^* = m_3^* + \delta_t \left(1 - \frac{s_3}{2}\right) \partial_{t_0}(Tv_x), \tag{48}$$

$$m_{5,\text{new}}^* = m_5^* + \delta_t \left(1 - \frac{s_5}{2}\right) \partial_{t_0}(Tv_y), \tag{49}$$

where $m_3^*$ and $m_5^*$ are given by Eq. (45). Meanwhile, according to the Chapman-Enskog analysis, we can find that (see Eqs. (A14) and (A15) in the Appendix)

$$-\partial_{t_0}(Tv_x) + \frac{1}{3}\partial_x(m_1^{eq} + m_2^{eq}) = -s_4 m_4^{(1)}, \tag{50}$$

$$-\partial_{t_0}(Tv_y) + \frac{1}{3}\partial_y(m_1^{eq} + m_2^{eq}) = -s_6 m_6^{(1)}. \tag{51}$$

Setting $m_1^{eq} + m_2^{eq} = 0$, the following relations can be obtained:

$$\partial_{t_0}(Tv_x) = s_4 m_4^{(1)}, \quad \partial_{t_0}(Tv_y) = s_6 m_6^{(1)}, \tag{52}$$

which means that $\partial_{t_0}(Tv_x)$ and $\partial_{t_0}(Tv_y)$ can be evaluated from $m_4^{(1)}$ and $m_6^{(1)}$, respectively. According to the setting of $m_1^{eq} + m_2^{eq} = 0$, the equilibria $\mathbf{m}^{eq}$ can be changed from Eq. (44) to

$$\mathbf{m}^{eq} = T(1, -2, 2, v_x, -v_x, v_y, -v_y, 0, 0)^{\mathrm{T}}. \tag{53}$$

The above equilibria can also be found in Ref. [37]. Using Eq. (52), the modifications given by Eqs. (48) and (49) can be rewritten as follows:

$$m^*_{3,\,\mathrm{new}} = m^*_3 + \delta_t\left(1 - \frac{s_3}{2}\right) s_4 m^{(1)}_4, \tag{54}$$

$$m^*_{5,\,\mathrm{new}} = m^*_5 + \delta_t\left(1 - \frac{s_5}{2}\right) s_6 m^{(1)}_6, \tag{55}$$

where the non-equilibrium parts $m^{(1)}_4$ and $m^{(1)}_6$ are calculated through $\delta_t \mathbf{m}^{(1)} \approx \mathbf{m} - \mathbf{m}^{eq}$. The Chapman-Enskog analysis in the Appendix shows that the target temperature equation can be correctly recovered. The idea of using the non-equilibrium parts of certain components in the moment space to adjust the macroscopic equations was introduced by Zheng *et al.* in Ref. [16], where they modified the collision processes of a D2Q17 MRT-LB model to achieve a consistent viscosity in the macroscopic momentum and energy equations. Similar treatments can also be found in the studies of Li *et al.* [12] and Huang and Wu [37].

To sum up, Eqs. (45), (46), (47), and (53) together with Eqs. (54) and (55) constitute the improved thermal LB model for liquid-vapor phase change. In numerical implementations, $\partial_t \phi$ in Eq. (46) is approximately calculated with $\partial_t \phi \approx [\phi(t) - \phi(t-\delta_t)]/\delta_t$ [34]. The isotropic difference schemes (see Eqs. (73) and (74) in Ref. [7]) are applied to the spatial gradients and the Laplacian of $T$ in the source term. For the flow field, an improved pseudopotential multiphase LB model proposed by Li *et al.* is employed (see Refs. [13,29] for details). The coupling between the multiphase LB model for flow field and the present thermal LB model for temperature field is established via the non-ideal equation of state, and we adopt the Peng-Robinson equation of state following the work of Ref. [48]:

$$p_{\mathrm{EOS}} = \frac{\rho RT}{1-b\rho} - \frac{a\vartheta(T)\rho^2}{1+2b\rho - b^2\rho^2}, \tag{56}$$

where $\vartheta(T) = \left[1 + (0.37464 + 1.54226\omega - 0.26992\omega^2)(1 - \sqrt{T/T_c})\right]^2$, $a = 0.45724 R^2 T_c^2/p_c$, and $b = 0.0778 RT_c/p_c$. The parameter $\omega = 0.344$ is the acentric factor and $T_c$ is the critical temperature, which can be obtained from the formulations of $a$ and $b$. In the present study, the saturation temperature of the system is chosen as $T_{\mathrm{sat}} = 0.86 T_c$. According to Ref. [48] and the relationship between $a$ and the interface thickness [13], we utilize $a = 3/49$, $b = 2/21$, and $R = 1$.

## IV. Numerical simulations

In this section, numerical simulations are carried out to validate the capability of the improved thermal model for simulating liquid-vapor phase change. For comparison, a compromised model is established, which is the same as the improved model except that no treatments are applied to eliminate the error term $\partial_{t_0}(T\mathbf{v})$ and the discrete effect of the source term (i.e., the error term $\partial_{t_0}\phi$). Hence, the effect of the replacement of $\nabla \cdot (\lambda \nabla T)/\rho c_V$ with $\nabla \cdot (\chi \nabla T)$ can be identified by a comparison of the numerical results between the Gong-Cheng and the compromised models. Meanwhile, the effect of the error terms can be identified by comparing the numerical results of the compromised model with those of the improved model. For different thermal models and a finite-difference scheme mentioned below, the flow simulation is fixed at using the aforementioned improved pseudopotential multiphase LB model so as to identify the performances of different solvers for the temperature equation.

### A. Droplet evaporation

First, the well-known D2 law for droplet evaporation is considered, which predicts that the square of the droplet diameter changes linearly over time [18,49]. This law is established based on the following conditions: the liquid and vapor phases are quasi-steady, the evaporation occurs in an environment with negligible viscous heat dissipation and no buoyancy, and the thermophysical properties (e.g., $c_V$ and $\lambda$) are constant. The simulations are carried out in a square domain with a grid size of $N_x \times N_y = 200 \times 200$ (lattice unit). Initially, a droplet with a diameter of $D_0 = 60$ is located in the center of the computational domain.

According to the requirement of the D2 law, no buoyant force is employed and the thermal conductivity is chosen to be constant: $\lambda = 2/3$ (lattice unit). Then the term $\nabla \cdot (\lambda \nabla T)/\rho c_V$ in Eq. (5) reduces to $\lambda \nabla^2 T/\rho c_V$. At the initial state, the temperature of the droplet is set to its saturation temperature, while a temperature $T_g$ is applied to the surrounding vapor of the droplet and the superheat $\Delta T = T_g - T_{sat}$ is chosen as $0.14T_c$. The droplet evaporation is driven by the temperature gradient at the liquid-vapor interface. At the boundaries, a constant temperature condition is employed ($T = T_g$). The relaxation parameters $s_3$ and $s_5$ are set to 1.0, which corresponds to $k = c_s^2 \delta_t / 2$ in Eq. (28). For the Gong-Cheng model, the relaxation time $\tau_g$ is given by $\tau_g = \lambda / (\rho c_s^2 c_V \delta_t) + 0.5$. The specific heat at constant volume is chosen as $c_V = 5$ and the kinematic viscosity is taken as $v = 0.1$ in the computational domain.

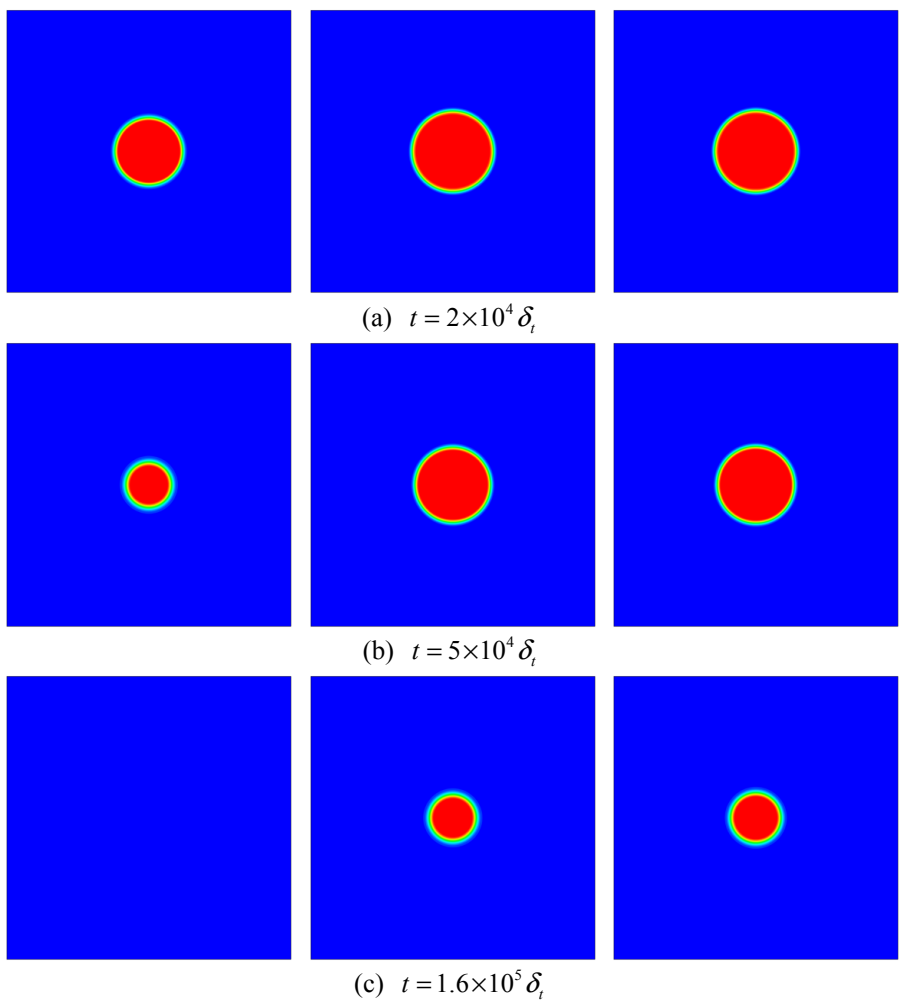

(a) $t = 2 \times 10^4 \delta_t$

(b) $t = 5 \times 10^4 \delta_t$

(c) $t = 1.6 \times 10^5 \delta_t$

**Fig. 1** Validation of the D2 law. Snapshots of the density contours obtained by the Gong-Cheng model (left), the compromised model (middle), and the improved model (right).

The snapshots of the density contours obtained by the Gong-Cheng model, the compromised model, and the improved model are shown in Fig. 1. The variation of $(D/D_0)^2$ with time is displayed in Fig. 2. For comparison, the available data in Ref. [50], which were obtained using a finite-difference scheme to solve the temperature equation (5) are also shown in Fig. 2. The figures show that the evaporation process predicted by the Gong-Cheng model is much faster than those predicted by the compromised model and the improved model. Moreover, Fig. 2 clearly shows that the numerical results of the Gong-Cheng model do not obey the D2 law (the square of the droplet diameter should change linearly over time), while the linear relationship can be observed in the results of the compromised model and the improved model. Furthermore, from Fig. 2 it can be seen that the numerical results given by the improved model are in excellent agreement with the date in Ref. [50].

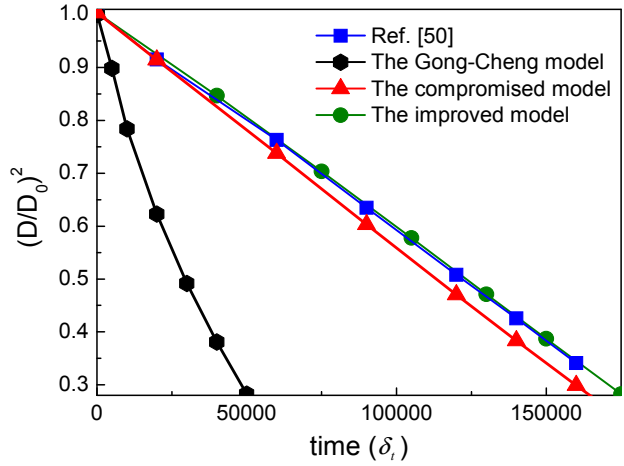

**Fig. 2** Validation of the D2 law. Comparison of the numerical results given by the Gong-Cheng model, the compromised model, and the improved model with the data in Ref. [50], which were obtained by a finite-difference scheme [29].

As mentioned earlier, the influence of the replacement of $\nabla \cdot (\lambda \nabla T)/\rho c_V$ with $\nabla \cdot (\chi \nabla T)$ can be identified by comparing the numerical results of the Gong-Cheng model with those of the compromised model. In Fig. 2 the severe deviations between the results of these two models indicate that such a treatment greatly affects the numerical results. These deviations are expected since evaporation is a type of vaporization that takes place at the surface of a liquid and in LB simulations the density varies remarkably within the liquid-vapor interface, which is usually a diffuse interface with a thickness of 4−5 lattices. Obviously, the aforementioned replacement is invalid within the liquid-vapor interface. Furthermore, from Fig. 2 we can observe some visible differences between the results of the compromised model and those of the improved model, which means that the error terms also yield considerable numerical errors.

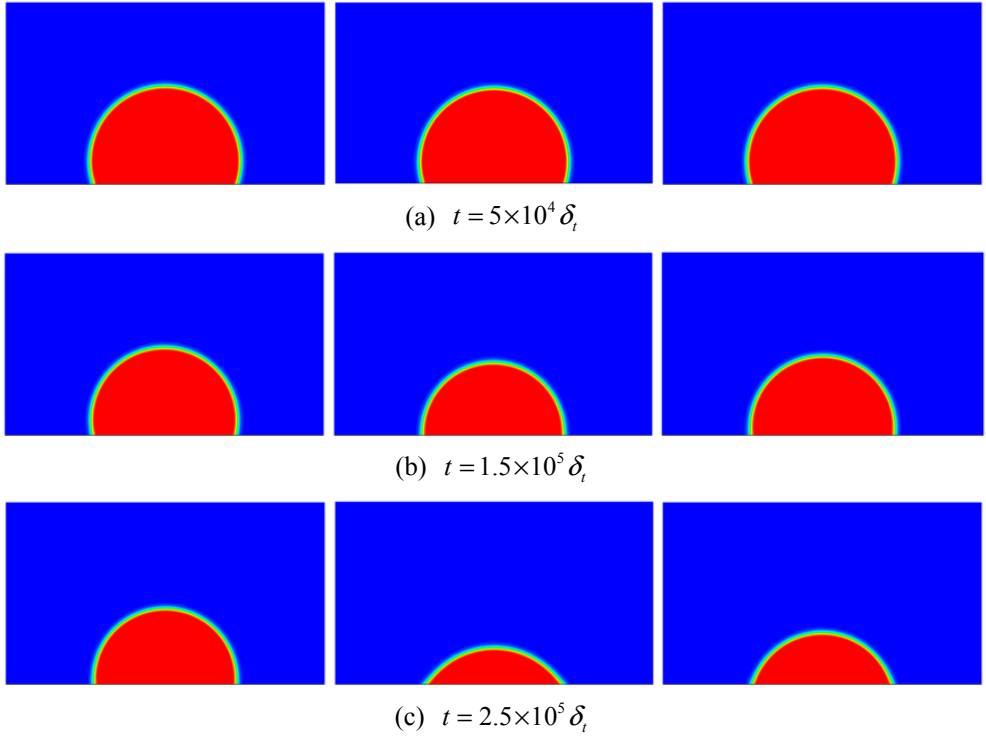

(a) $t = 5\times 10^4 \delta_t$

(b) $t = 1.5\times 10^5 \delta_t$

(c) $t = 2.5\times 10^5 \delta_t$

**Fig. 3** Droplet evaporation on a solid surface. Snapshots of the density contours obtained by the Gong-Cheng model (left), the compromised model (middle), and the improved model (right). The displayed domain is $x \in [50, 250]$ and $y \in [0, 115]$.

To further illustrate the above points, the droplet evaporation on a solid surface is also considered. In the above test, the thermal conductivity is chosen to be constant according to the requirement of the D2 law. In the present test, the thermal conductivity is taken as $\lambda = \rho c_V \chi$ with $\chi = 0.08$. Then $\nabla \cdot (\lambda \nabla T)$ should be treated as $\nabla \cdot (\lambda \nabla T) = \lambda \nabla^2 T + \nabla \lambda \cdot \nabla T$. The simulations are performed in a rectangular domain with a grid size of $N_x \times N_y = 300 \times 150$. A droplet with a radius of $r = 40$ is initially placed on the center of the bottom surface. The kinematic viscosity and the specific heat at constant volume are still set to $\nu = 0.1$ and $c_V = 5$, respectively. The temperature of the bottom surface is fixed at $T_w = 0.875 T_c$. The Zou-He boundary scheme [51] is applied to the solid surface and the open boundary condition is employed at the top boundary. The periodic boundary condition is utilized in the $x$-direction. The first 20000 steps of the simulations are carried out without evaporation so that the droplet can reach its equilibrium state. The equilibrium contact angle is taken as $\theta \approx 108^\circ$.

The thermal LB models are added after $t = 2\times 10^4 \delta_t$ and the contact angle hysteresis [52] is taken into consideration with a hysteresis window of $\left(0^\circ, 180^\circ\right)$

Figure 3 displays the snapshots of the density contours obtained by the Gong-Cheng model, the compromised model, and the improved model. Owing to the contact angle hysteresis, the droplet evaporates in the constant contact radius (CCR) mode, namely the contact angle decreases whereas the contact line is pinned on the solid surface. Figure 3 shows that in the present test the evaporation process predicted by the Gong-Cheng model is slower than those predicted by the compromised model and the improved model, which is found to be related to the choose of a variable thermal conductivity in the present test. When a constant $\lambda$ is applied in the present test, the evaporation process given by the Gong-Cheng model is faster than those given by the other two models.

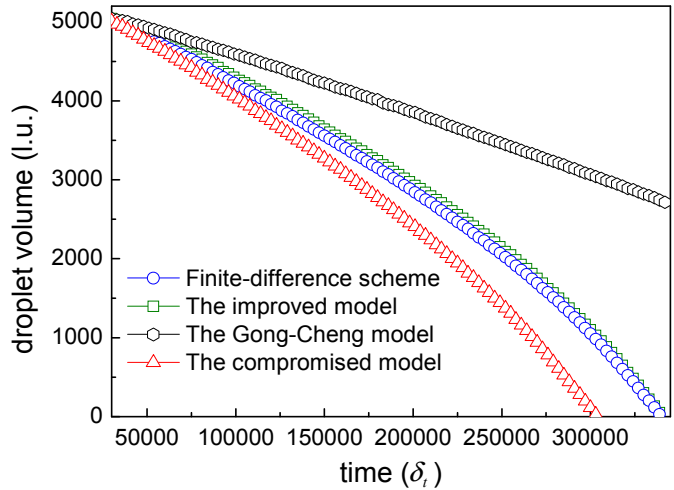

**Fig. 4** Droplet evaporation on a solid surface. Comparison of the numerical results obtained by the Gong-Cheng model, the compromised model, the improved model, and a finite-difference scheme [29].

The variation of the droplet volume with time is shown in Fig. 4, where l.u. represent lattice units. For comparison, the numerical results obtained by a finite-difference scheme for solving the temperature equation [29] are also shown in Fig. 4. From the figure we can see that the numerical results of the improved model agree well with those obtained by the finite-difference scheme. Similarly, Fig. 4 also shows that there are significant deviations between the numerical results of the Gong-Cheng model and those of the compromised model, which arise from the replacement of $\nabla \cdot (\lambda \nabla T)/\rho c_V$ with $\nabla \cdot (\chi \nabla T)$. Moreover, considerable deviations, which are caused by the error terms, can be observed between the compromised model and the improved model.

## B. Bubble nucleation and departure

In this subsection, numerical simulations are performed for bubble nucleation and departure involved in nucleate boiling. Our simulations are carried out in a rectangular domain with a grid size of $N_x \times N_y = 150 \times 300$. The kinematic viscosity, the specific heat at constant volume, the saturation temperature, and the relaxation parameters are the same as those used in the previous subsection. The thermal conductively is taken as $\lambda = \rho c_V \chi$ with $\chi = 0.06$. The initial setting of the computational domain is a liquid ($0 \leq y \leq 0.5 N_y$) below its vapor, and the temperature in the domain is set to $T_{sat}$. The temperature of the bottom wall is fixed at $T_{sat}$ except that a high temperature $T_h = 1.25 T_c$ is applied to the central three grids of the wall. The equilibrium contact angle is taken as $\theta \approx 45°$. The periodic boundary condition is applied to $x$-direction. The buoyant force is given by $\mathbf{F}_b = (\rho - \rho_{ave}) \mathbf{g}$, where $\mathbf{g} = (0, -g)$ is the gravitational acceleration and $\rho_{ave}$ is the average density in the domain.

The snapshots of the density contours obtained by the improved model, the compromised model, and the Gong-Cheng model with the gravitational acceleration $g = 1.5 \times 10^{-5}$ are shown in Fig. 5. From the results of the improved model, it can be seen that a bubble has nucleated at $t = 2000 \delta_t$ owing to the high temperature at the center of the bottom wall. The vapor bubble gradually grows until its diameter reaches the departure diameter. Then the bubble detaches from the solid wall, which can be seen from the third snapshot of the numerical results of the improved model. After the detachment, a tiny attached bubble remains on the bottom wall, which repeats the behavior of the first bubble. Similar to the previous two tests, the present test also shows that the numerical results of the Gong-Cheng model significantly deviate from those of the other two models and some visible differences can be observed between the numerical results of the compromised model and those of the improved model.

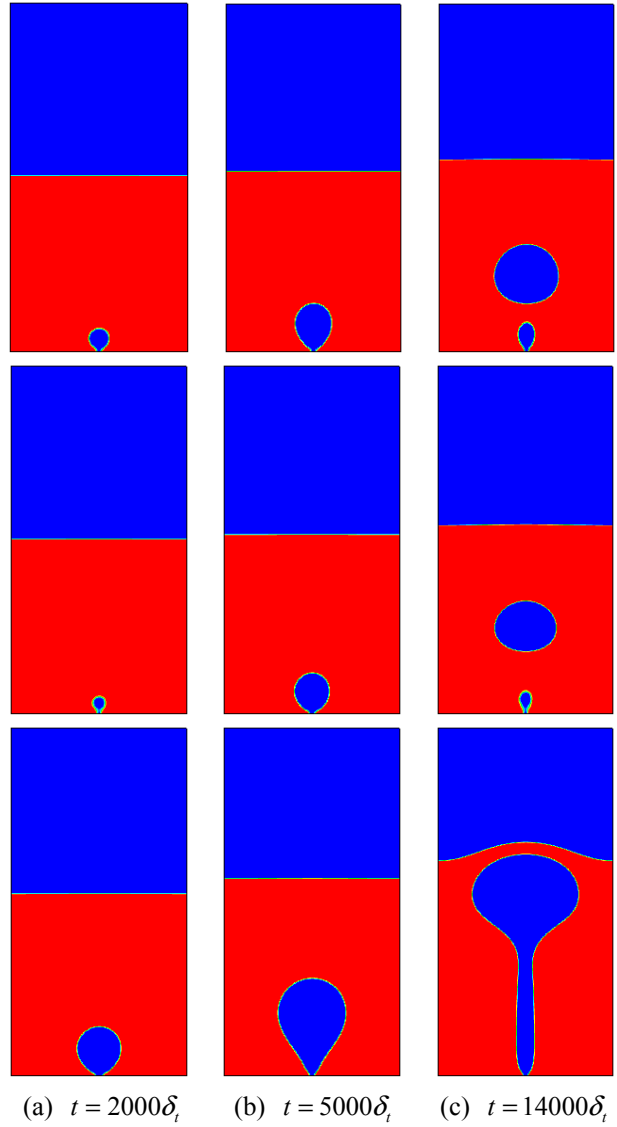

(a) $t = 2000\delta_t$   (b) $t = 5000\delta_t$   (c) $t = 14000\delta_t$

**Fig. 5** Simulation of bubble nucleation and departure ($g = 1.5 \times 10^{-5}$). Snapshots of the density contours obtained by the improved model (top), the compromised model (middle), and the Gong-Cheng model (bottom).

Figure 6 displays the snapshots of the density contours obtained by the improved model, the compromised model, and the Gong-Cheng model with the gravitational acceleration $g = 2.5 \times 10^{-5}$. Similarly, the numerical results of the Gong-Cheng model are remarkably different from those of the other two models, further confirming that the replacement of $\nabla \cdot (\lambda \nabla T)/\rho c_V$ with $\nabla \cdot (\chi \nabla T)$ results in significant numerical errors. Moreover, a comparison of the numerical results in Figs. 5 and 6 between the compromised model and the improved model shows that the error terms $\partial_{t_0}(T\mathbf{v})$ and $\partial_{t_0}\phi$ affect bubble growth and the bubble departure diameter. Meanwhile, the numerical results of the improved model in Figs. 5 and 6 show that the bubble departure diameter decreases with the increase of the gravitational acceleration $g$. Quantitatively, the bubble departure diameter obtained by the improved model is plotted in Fig. 7 against the gravitational acceleration $g$, where the symbols represent the numerical results while the solid line represents the results of $0.209g^{-0.5}$. The figure illustrates that the bubble departure diameter predicted by the improved model is proportional to $g^{-0.5}$, which is consistent with the correlations in the literature [53].

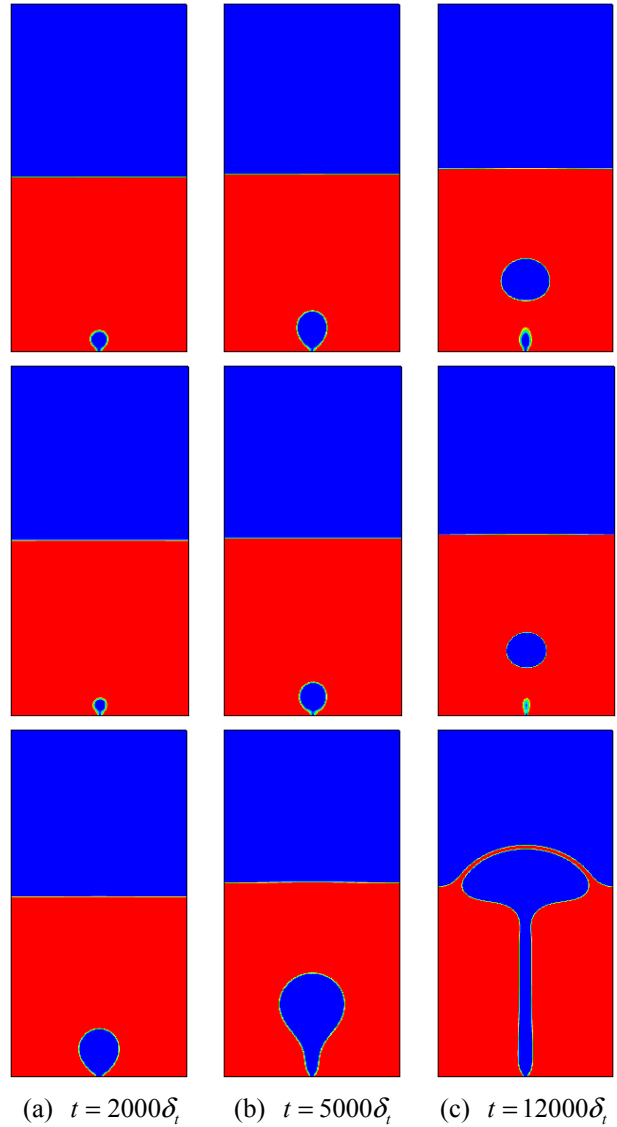

(a) $t = 2000\delta_t$  (b) $t = 5000\delta_t$  (c) $t = 12000\delta_t$

**Fig. 6** Simulation of bubble nucleation and departure ($g = 2.5 \times 10^{-5}$). Snapshots of the density contours obtained by the improved model (top), the compromised model (middle), and the Gong-Cheng model (bottom).

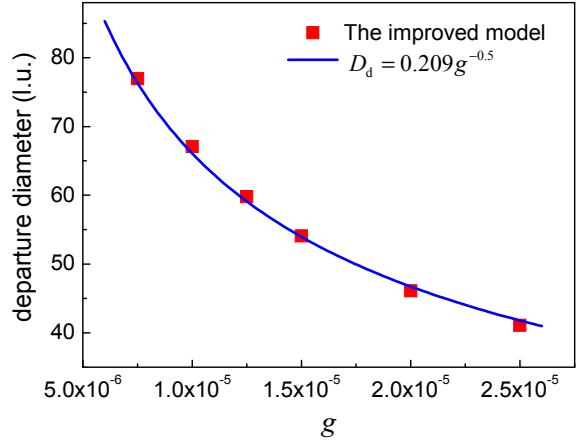

**Fig. 7** Simulation of bubble nucleation and departure. The bubble departure diameter predicted by the improved model. The squares represent the numerical results obtained by the improved model and the solid line denotes the results given by $D_\text{d} = 0.209 g^{-0.5}$.

## V. Conclusions

In this paper, we have presented an improved thermal LB model for simulating liquid-vapor phase change. The Chapman-Enskog analysis has been performed for the Gong-Cheng model, which shows that the term $\nabla \cdot (\lambda \nabla T)/\rho c_V$ in the target temperature equation was replaced by $\nabla \cdot (\chi \nabla T)$ in the model and some error terms exist in the recovered macroscopic temperature equation. Theoretical analyses have been provided about removing the error terms within the framework of the BGK collision operator. The improved thermal LB model was constructed based on the MRT collision operator. The error terms $\partial_{t_0}(T\mathbf{v}) + \nabla \cdot (T\mathbf{vv})$ as well as the discrete effect of the source term have been eliminated in the improved model.

Numerical simulations have been carried out for droplet evaporation and bubble nucleation and departure involved in nucleate boiling to validate the capability of the improved model. For comparison, a compromised model was established, which is the same as the improved model except that no treatments are applied to eliminate the error terms $\partial_{t_0}(T\mathbf{v})$ and $\partial_{t_0}\phi$. By comparing the numerical results of the Gong-Cheng model with those of the compromised model, it is demonstrated that the replacement of $\nabla \cdot (\lambda \nabla T)/\rho c_V$ with $\nabla \cdot (\chi \nabla T)$ yields significant numerical errors. Moreover, by comparing the numerical results of the compromised model with those of the improved model, it is found that the numerical errors caused by the error terms are non-negligible. We believe that the theoretical analyses as well as the numerical results in the present paper are useful for clarifying some critical issues and the present study would be helpful for general readers to better understand the thermal LB models for liquid-vapor phase change.

## Acknowledgments

This work was supported by the National Natural Science Foundation of China (No. 51506227) and the Foundation for the Author of National Excellent Doctoral Dissertation of China (No. 201439).

## Appendix: The Chapman-Enskog analysis of the improved thermal MRT-LB model

The Taylor series expansion of Eq. (42) yields

$$\delta_t (\partial_t + \mathbf{e}_\alpha \cdot \nabla) g_\alpha + \frac{\delta_t^2}{2} (\partial_t + \mathbf{e}_\alpha \cdot \nabla)^2 g_\alpha + \cdots = -\bar{\Lambda}_{\alpha\beta} \left( g_\beta - g_\beta^{eq} \right)|_{(\mathbf{x}, t)} + \delta_t S'_\alpha (\mathbf{x}, t). \quad (A1)$$

Using the multi-scale expansions, Eq. (A1) can be rewritten in the consecutive orders of $\delta_t$ as follows:

$$O(\delta_t): \left( \partial_{t_0} + \mathbf{e}_\alpha \cdot \nabla \right) g_\alpha^{eq} = -\bar{\Lambda}_{\alpha\beta} g_\beta^{(1)}|_{(\mathbf{x}, t)} + S'^{(0)}_\alpha, \quad (A2)$$

$$O(\delta_t^2): \partial_{t_1} g_\alpha^{eq} + \left( \partial_{t_0} + \mathbf{e}_\alpha \cdot \nabla \right) g_\alpha^{(1)} + \frac{1}{2} \left( \partial_{t_0} + \mathbf{e}_\alpha \cdot \nabla \right)^2 g_\alpha^{eq} = -\bar{\Lambda}_{\alpha\beta} g_\beta^{(2)}|_{(\mathbf{x}, t)} + S'^{(1)}_\alpha. \quad (A3)$$

Multiplying Eqs. (A2) and (A3) with the transformation matrix $\mathbf{M}$ lead to the following equations:

$$O(\delta_t): \mathbf{D}_0 \mathbf{m}^{eq} = -\mathbf{\Lambda} \mathbf{m}^{(1)} + \mathbf{S}^{(0)}, \quad (A4)$$

$$O(\delta_t^2): \partial_{t_1} \mathbf{m}^{eq} + \mathbf{D}_0 \mathbf{m}^{(1)} + \frac{1}{2} \mathbf{D}_0^2 \mathbf{m}^{eq} = -\mathbf{\Lambda} \mathbf{m}^{(2)} + \mathbf{S}^{(1)}, \quad (A5)$$

where $\mathbf{D}_0 = \partial_{t_0} \mathbf{I} + \bar{\mathbf{C}} \cdot \nabla$, in which $\bar{\mathbf{C}} \cdot \nabla = \bar{C}_x \partial_x + \bar{C}_y \partial_y$, $\mathbf{S}^{(0)} = (\phi, 0, 0, 0, 0, 0, 0, 0, 0)^{\mathrm{T}}$, and $\mathbf{S}^{(1)} = (0.5 \partial_{t_0} \phi, 0, 0, 0, 0, 0, 0, 0, 0)^{\mathrm{T}}$. The detailed forms of $\bar{C}_x$ and $\bar{C}_y$ for the D2Q9 lattice can be found in Ref. [54]. Substituting Eq. (A4) into Eq. (A5), we can obtain

$$\partial_{t_1} \mathbf{m}^{eq} + \mathbf{D}_0 \left( \mathbf{I} - \frac{\mathbf{\Lambda}}{2} \right) \mathbf{m}^{(1)} + \frac{1}{2} \mathbf{D}_0 \mathbf{S}^{(0)} = -\mathbf{\Lambda} \mathbf{m}^{(2)} + \mathbf{S}^{(1)}. \tag{A6}$$

According to Eq. (A4), we have

$$\partial_{t_0} T + \partial_x (T v_x) + \partial_y (T v_y) = \phi, \tag{A7}$$

$$\partial_{t_0} (T v_x) + \partial_x (c_s^2 T) = -s_3 m_3^{(1)}, \tag{A8}$$

$$\partial_{t_0} (T v_y) + \partial_y (c_s^2 T) = -s_5 m_5^{(1)}. \tag{A9}$$

From Eq. (A6), we can obtain

$$\partial_{t_1} T + \partial_x \left[ \left(1 - \frac{s_3}{2}\right) m_3^{(1)} \right] + \partial_y \left[ \left(1 - \frac{s_5}{2}\right) m_5^{(1)} \right] + \frac{1}{2} \partial_{t_0} \phi = \frac{1}{2} \partial_{t_0} \phi, \tag{A10}$$

which further yields

$$\partial_{t_1} T + \partial_x \left[ \left(1 - \frac{s_3}{2}\right) m_3^{(1)} \right] + \partial_y \left[ \left(1 - \frac{s_5}{2}\right) m_5^{(1)} \right] = 0. \tag{A11}$$

With the aid of Eqs. (A8) and (A9) and setting $s_5 = s_3$, Eq. (A11) can be written as

$$\partial_{t_1} T = \partial_x \left( \eta c_s^2 \partial_x T \right) + \partial_y \left( \eta c_s^2 \partial_y T \right) + \partial_x \left[ \eta \partial_{t_0} (T v_x) \right] + \partial_y \left[ \eta \partial_{t_0} (T v_y) \right]$$
$$= \nabla \cdot \left( \eta c_s^2 \nabla T \right) + \nabla \cdot \left[ \eta \partial_{t_0} (T \mathbf{v}) \right], \tag{A12}$$

where $\eta$ is given by

$$\eta = \left( \frac{1}{s_3} - \frac{1}{2} \right) = \left( \frac{1}{s_5} - \frac{1}{2} \right). \tag{A13}$$

Meanwhile, according to Eq. (A4), we can obtain

$$-\partial_{t_0} (T v_x) + \frac{1}{3} \partial_x \left( m_1^{eq} + m_2^{eq} \right) = -s_4 m_4^{(1)}, \tag{A14}$$

$$-\partial_{t_0} (T v_y) + \frac{1}{3} \partial_y \left( m_1^{eq} + m_2^{eq} \right) = -s_6 m_6^{(1)}. \tag{A15}$$

When the equilibria $\mathbf{m}^{eq}$ are defined by Eq. (44), $m_1^{eq} + m_2^{eq} = -T$. However, when the equilibria $\mathbf{m}^{eq}$ are given by Eq. (53), we have $m_1^{eq} + m_2^{eq} = 0$. Then the following relations can be obtained:

$$\partial_{t_0} (T v_x) = s_4 m_4^{(1)}, \tag{A16}$$

$$\partial_{t_0} (T v_y) = s_6 m_6^{(1)}. \tag{A17}$$

In other words, $\partial_{t_0} (T v_x)$ and $\partial_{t_0} (T v_y)$ can be evaluated with $m_4^{(1)}$ and $m_6^{(1)}$, respectively.

With the modifications given by Eqs. (54) and (55), the following equations can be obtained:

$$\partial_{t_0}(Tv_x) + \partial_x(c_s^2 T) = -s_3 m_3^{(1)} + \left(1 - \frac{s_3}{2}\right) s_4 m_4^{(1)}, \tag{A18}$$

$$\partial_{t_0}(Tv_y) + \partial_y(c_s^2 T) = -s_5 m_5^{(1)} + \left(1 - \frac{s_5}{2}\right) s_6 m_6^{(1)}. \tag{A19}$$

Similarly, Eq. (A10) will become

$$\partial_{t_1} T + \partial_x\left[\left(1 - \frac{s_3}{2}\right) m_3^{(1)}\right] + \partial_y\left[\left(1 - \frac{s_5}{2}\right) m_5^{(1)}\right] + \frac{1}{2}\partial_x\left[\left(1 - \frac{s_3}{2}\right) s_4 m_4^{(1)}\right] + \frac{1}{2}\partial_y\left[\left(1 - \frac{s_5}{2}\right) s_6 m_6^{(1)}\right] = 0. \tag{A20}$$

Substituting Eqs. (A18) and (A19) into Eq. (A20) and using Eqs. (A16) and (A17), we can obtain

$$\partial_{t_1} T = \partial_x\left[\left(\frac{1}{s_3} - \frac{1}{2}\right) c_s^2 \partial_x T\right] + \partial_y\left[\left(\frac{1}{s_5} - \frac{1}{2}\right) c_s^2 \partial_y T\right]. \tag{A21}$$

Setting $s_5 = s_3$ and combining Eq. (A21) with Eq. (A7) through $\partial_t = \partial_{t_0} + \delta_t \partial_{t_1}$, we have

$$\partial_t T + \nabla \cdot (\mathbf{v} T) = \nabla \cdot (k \nabla T) + \phi, \tag{A22}$$

where $k = \eta c_s^2 \delta_t$ and $\eta$ is given by Eq. (A13).